# Field-dependent surface impedance tensor in amorphous wires with two types of magnetic anisotropy: helical and circumferential

D. P. Makhnovskiy, L. V. Panina, and D. J. Mapps

Department of Communication, Electronic and Electrical Engineering, University of Plymouth, Drake Circus, Plymouth, Devon PL4 8AA, United Kingdom

**Abstract**

This paper concerns the theoretical and experimental investigation of the magneto-impedance (MI) effect in amorphous wires in terms of the surface impedance tensor $\hat{V}$. Physical concepts of MI and problems of significant practical importance are discussed using the results obtained. The theoretical analysis is based on employing the asymptotic-series expansion method of solving the Maxwell equations for a ferromagnetic wire with an ac permeability tensor of a general form associated with magnetisation rotation. The magnetic structure-dependent impedance tensor $\hat{V}$ is calculated for any frequency and external magnetic field, and is not restricted to the case when only strong skin-effect is present. This approach allows us to develop a rigorous quantitative analysis of MI characteristics in wires, depending on the type of magnetic anisotropy, the magnitude of dc bias current, and an excitation method. The theoretical model has been tested by comparing the obtained results with experiment. For the sake of an adequate comparison, the full tensor $\hat{V}$ is measured in CoFeSiB and CoSiB amorphous wires having a circumferential and helical anisotropy, respectively, by determining the $S_{21}$ parameter. In cases, when the rotational dynamics is responsible for the impedance behaviour, there is a reasonable agreement between the experimental and theoretical results. Such effects as the ac biased asymmetrical MI in wires with a circumferential anisotropy, the transformation in MI behaviour caused by a dc current (from that having a symmetric hysteresis to an asymmetric anhysteretic one) in wires with a helical anisotropy are discussed.

## I. Introduction

This paper addresses the magnetic-structure dependent impedance analysis in amorphous magnetic wires with a helical (or circumferential) anisotropy, including such practically important phenomena as the role of a magnetic anisotropy and a dc bias current in controlling magneto-impedance (MI) characteristics, asymmetrical MI and the role of the off-diagonal impedance in asymmetrical MI. We have also carried out an experimental investigation of these effects to be able to demonstrate the consistency between the theoretical and experimental results. Since the discovery of the MI effect in 1993,[1] it has received much attention due to its importance for developing new generation micro-magnetic sensors of high performance.[1-4] However, most of the theoretical work is restricted to specific conditions not always consistent with the experiment. In certain cases, conflicting experimental results on MI in materials with a similar magnetic structure have been reported. This has occurred when different types of excitation have been used. This is particularly related to the case of a complicated magnetic configuration, as the case of a helical magnetisation in a ferromagnetic wire. Therefore, rigorous theoretical and experimental research of MI effects accounting for specific magnetic structures and excitation methods remains to be of a considerable interest and importance.

In general, the MI effect involves a very large and sensitive change in the voltage measured across a ferromagnetic specimen with a well-defined transverse magnetic anisotropy, carrying a high frequency current and subjected to a dc magnetic field. For example, in the case of 30 μm diameter amorphous wires of a composition $(Co_{0.94}Fe_{0.06})_{72.5}Si_{12.5}B_{15}$ (having a circumferential anisotropy), the voltage (or impedance) change can be as much as 10 to 50%/Oe at frequencies of several megahertz.[5-8] Considering MI as a change in a complex resistance, it has a direct analogy with giant magneto resistance (GMR). In earlier work there was an attempt to regard it as a new ac GMR effect, explained in terms of ac quantum magneto-transport.[9,10] However, this approach failed to explain a very large nominal change in the measured voltage as a change in true resistance, as well as the voltage dependence with frequency in the presence of a dc magnetic field. It soon became clear that the effect has an electrodynamic origin owing to the redistribution of the ac current density under the application of the dc magnetic field. In the original theoretical work on MI [5,6] the current density has been calculated with the assumption that the variable magnetic properties can be described in terms of a total permeability having a scalar or quasi-diagonal form. This allows the impedance of a magnetic object (and the voltage induced across it by the ac current $i = i_0 \exp(-j\omega t)$) to be found essentially in the same way as in the case of a non-magnetic material.[11] In this approach, the voltage response $V$ is of the form

$$V = Z(a/\delta_m)i \qquad (1)$$

where the impedance $Z$ is calculated as a function of a skin depth

$$\delta_m = c/\sqrt{2\pi\sigma\omega\mu_t} \qquad (2)$$

Here $c$ is the velocity of light, $\sigma$ is the conductivity and $\mu_t$ is the effective transverse permeability (with respect to the current flow), $2a$ is a characteristic cross-section size. If the skin effect is strong $a/\delta_m \gg 1$, the impedance is inversely proportional to the skin depth, therefore, the magnetic-field dependence of the transverse permeability controls the voltage behaviour. This simple consideration has provided a qualitative understanding of the MI behaviour, and in certain cases equations (1), (2) have given a reasonable agreement with the experimental results. A good example is the MI effect in a Co-based amorphous wire. A tensile stress from quenching (and enhanced by tension annealing) coupled with the negative magnetostriction results in a circumferential anisotropy and a corresponding left and right handed alternative circular domain structure.[12,13] The ac current passing through the wire induces an easy-axis magnetic field which moves the circular domain walls so that they nearly cross the entire wire. The circular magnetisation is very sensitive to the axial dc magnetic field which is a hard axis field. The ac permeability associated with this process is circumferential and corresponds to $\mu_t$ introduced in (2). Substituting in equations (1),(2) this circular permeability accounting for the field-dependence and the frequency dispersion due to the local domain wall damping gives a very good agreement with the experimental MI spectra for frequencies lower than the characteristic frequency of the domain wall relaxation ($\sim 1-10$ MHz for 30 micron diameter wires).[5,6] Typically, the rotational relaxation is a faster process, and for higher frequencies the magnetisation rotation dynamics dominates. The rotational permeability has an essential tensor form, which makes it difficult to use equations (1), (2) for higher frequencies: the difference between the experimental and theoretical results becomes quite considerable.

Further experiments on MI have resulted in the discovery of such phenomena as asymmetrical or bi-stable MI in twisted (or torsion annealed) amorphous wires,[14-18] asymmetrical MI in annealed amorphous ribbons [19,20] and in films with crossed anisotropy,[21] and the effect of an ac bias field producing asymmetrical voltage response in systems having no magnetic asymmetry in the dc magnetic configuration.[22,23] Regarding these phenomena, the approach based on equations (1),(2) can fail to provide even a qualitative explanation, especially in the case of the ac biased asymmetrical MI.

Another theoretical difficulty is related to the case of MI in the multilayered films consisting of two upper magnetic layers sandwiching a non-magnetic conductor. If the film width is smaller than a certain critical value, the magnetic flux leakage through the inner conductor becomes essential in the determining the impedance tensor of the total system. This effect is known to give a considerable contribution to a high frequency inductance of similar systems. The existing theoretical approach to this problem [24-27] does not account for the tensor form of the permeability, which is not correct for a practically important case of a transverse (or crossed) magnetic structure in the outer layers.[28]

Therefore, numerous experimental results on MI require a more realistic theory taking into account a specific tensor form of the ac permeability and impedance. In the present paper, a general approach to solving electrodynamic problems for ferromagnetic objects characterised by a given permeability tensor is proposed, which is based on the expansion of Maxwell's equations in asymptotic series. The characteristic parameter of these expansions can be chosen to be the ratio $\beta = a/\delta$ where $\delta = c/\sqrt{2\pi\sigma\omega}$ is the non-magnetic penetration depth ($\mu_t = 1$). Constructing the asymptotic serious for two limiting cases $\beta \gg 1$ and $\beta \ll 1$ and matching them in the intermediate region, the solution for ac field distribution becomes valid in the entire frequency (or dc magnetic field) range. For obtaining the asymptotic series in the case $\beta \gg 1$, a singular perturbation method is used, which is needed to describe the field distribution in the surface layer. For $\beta \ll 1$, a standard regular perturbation method can be employed. The asymptotic method for solving the problems of electrodynamics such as the impedance analysis in ferromagnetic conductors has been used for the first time in this case, although it has been known in such fields as heat transmission, diffusion and certain problems in optics.[29-31]

The method of asymptotic series is applied to the calculation of the surface impedance tensor in a magnetic wire having in general a helical magnetic anisotropy. The ac magnetisation is assumed to be related to the rotational process and is described by a tensor of a general form having 6 different components. Considering, that the wire is subjected to an ac current $i$ and an ac axial field $h_{ex}$, and its static magnetic structure can be modified by a dc axial magnetic field $H_{ex}$ and a dc current $I_b$ (see Fig. 1), a quantitative explanation of a number of high frequency MI effects in wires can be given. This includes the modification of MI characteristics under the effect of the dc current in a wire with a helical anisotropy. Without dc bias, the plots of impedance vs. $H_{ex}$

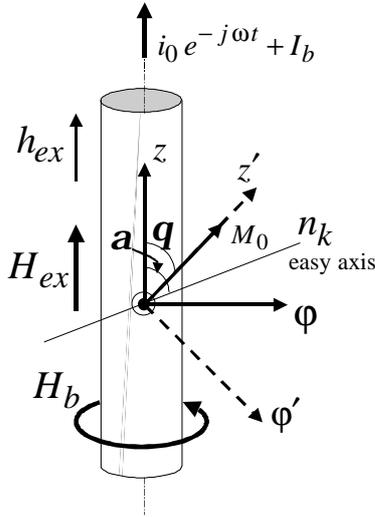

Fig. 1. Principle directions and quantities used

exhibit a symmetric hysteresis. With increasing the bias field, the hysteresis area shifts and shrinks, and finally disappears, resulting in highly sensitive asymmetric impedance plots.

To demonstrate consistency between theory and experiment, measurements of the impedance tensor in amorphous wires with both types of anisotropy have been made under proper conditions. A number of previously obtained results have been repeated here for the sake of an accurate comparison, since the MI behaviour depends significantly on the excitation conditions. In the case of a helical anisotropy, the full impedance tensor has been obtained for the first time. The role of the off-diagonal components of the impedance can be seen if the voltage response is measured in the external coil, or when the wire is subjected to the ac axial field. For example, in the presence of both $i$ and $h_{ex}$, the voltage measured across the MI element exhibits a strong asymmetry which is due to the contribution of the off-diagonal tensor component.

The approach of the surface impedance tensor (and a tensor permeability) to described the MI phenomena has been previously used in a number of works using certain simplifications. In the case of magnetic/metallic multilayers,[28,32] the edge effects have been ignored completely, and the film system is treated as having infinite dimensions in plane. The asymptotic methods developed in the present work for a wire geometry can be modified for a two-dimensional impedance analysis in the multilayer film which will be published elsewhere. Regarding the wire geometry, in Ref. 15 the asymmetrical MI has been considered for a low frequency limit ($b \ll 1$). The method used turns out to have a very slow convergence, which has provided a qualitative approach only. In Ref. 33, the off-diagonal impedance has been analysed for a wire with a circumferential anisotropy, which does not include the effect of a dc current to produce asymmetrical MI. Besides, the approximation of very high frequencies ($b \gg 1$) has been used, treating the wire as a plane object. It seems that this approximation is not consistent with the experimental case. Almost all
the experimental results on MI are obtained for 30-micron diameter amorphous wires having the resistance of 130 $\mu\Omega \cdot cm$. In this case, the condition $b \gg 1$ requires the frequency to be in the GHz range, whereas as the experiment is concerned with frequencies of 1–100 MHz. However, the range of frequencies and fields where this approximation is reasonable is much wider, as it has been proved in the present analysis. Calculating the higher-order terms in the expansion (in the parameter $1/b$), we have demonstrated that they contain a certain magnetic parameter $m_{ef}$ and the actual validity condition is $\sqrt{m_{ef}}\, b \gg 1$. It is worth noticing that it was not possible to obtain this important conclusion within the model used in Ref. 33, since calculating the impedance for a plane geometry gives only zero-degree terms and does not allow the next terms to be determined.

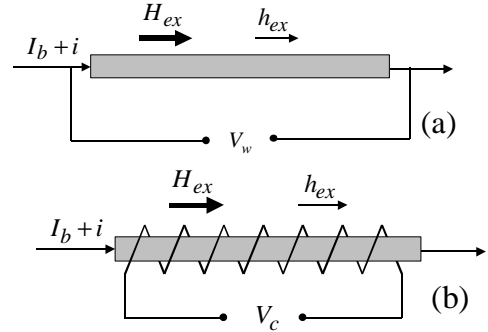

Fig. 2. Voltage response due to the ac excitation using current $i$ and field $h_{ex}$, measured across the wire in (a) and in the coil in (b).

The paper is organised as follows: Section II introduces the surface impedance tensor with relation to a certain ac excitation and voltage response measurement. Section III formulates the problem, presenting the linearised Maxwell's equations and the permeability tensor for the model under consideration. Sections IV and V give solutions in high ($b \gg 1$) and low ($b \ll 1$) frequency limits, respectively. Section VI concerns the numerical analysis of the behaviour of the impedance tensor in a single domain wire with two types of anisotropy (circumferential and helical). Finally, in Section VII the experimental results are given, demonstrating a very good agreement with the theory.

## II. Voltage response and surface impedance tensor

The GMI effect deals with a voltage response in a thin metallic magnetic material subjected to a high frequency excitation. In the case of a wire, it is reasonable to use an ac current $i$ and/or an ac axial field $h_{ex}$ as a source of excitation (see Fig. 2). The voltage is measured either across the wire ($V_w$) or in the coil ($V_c$) mounted on it. The value of $V_w$ is determined by considering the energy consumption in the wire

$$iV_w = \frac{c}{4\pi} \int_S (\mathbf{e}\times\mathbf{h}) \mathbf{ds} \qquad (3)$$

where the integration is performed along the wire surface, $\mathbf{e}$ and $\mathbf{h}$ are the ac electric and magnetic fields, $c$ is the velocity of light. The voltage $V_c$ is found by integrating $\mathbf{e}$ along the coil turns

$$V_c = \oint \mathbf{e}\,\mathbf{dl} \qquad (4)$$

As it follows from (3), (4), the induced voltage can be found by calculating the tangential components of the fields $\bar{\mathbf{e}}_t$, $\bar{\mathbf{h}}_t$ at the wire surface. Since it is assumed that the wavelength is larger than the sample size, the field distribution outside the sample corresponds to the static case. Then, the excitation method imposes the boundary conditions for the magnetic field $\bar{\mathbf{h}}_t$. Using the cylindrical co-ordinates ($z, \varphi, r$) with the axis z along the wire (see Fig. 1), the boundary conditions can be written as

$$h_\varphi(a) = \bar{h}_\varphi = 2i/ca, \qquad h_z(a) = h_{ex} \qquad (5)$$

where $a$ is the wire radius. The electric field $\bar{\mathbf{e}}_t$ is related to the magnetic field $\bar{\mathbf{h}}_t$ via the surface impedance tensor $\hat{\zeta}$

$$\bar{\mathbf{e}}_t = \hat{\zeta}(\bar{\mathbf{h}}_t \times \mathbf{n}) \qquad (6)$$

where $\mathbf{n}$ is a unit radial vector directed inside the wire. Comparing (3)-(6) it is seen that the impedance $\hat{\zeta}$ is the only characteristic describing the voltage response in the system excited by the external magnetic field $\mathbf{h}$ (of any origin). In ferromagnetic conductors, $\hat{\zeta}$ is a two-dimensional tensor even for the electrically isotropic case.

The present analysis is concerned with the calculation of the surface impedance tensor for a wire with a uniform static magnetisation having a helical orientation. In this case, the tensor $\hat{\zeta}$ is constant on the surface. Writing vector equation (6) in the co-ordinate representation, the components of $\hat{\zeta}$ can be determined as

$$\begin{aligned}\bar{e}_\varphi &= -\zeta_{\varphi\varphi}\,\bar{h}_z + \zeta_{\varphi z}\,\bar{h}_\varphi \\ \bar{e}_z &= -\zeta_{z\varphi}\,\bar{h}_z + \zeta_{zz}\,\bar{h}_\varphi \end{aligned} \qquad (7)$$

where $\zeta_{z\varphi} = \zeta_{\varphi z}$ because of symmetry. Substituting (7) to (3) and (4) gives the voltage responses

$$V_w = \bar{e}_z L = (\zeta_{zz}\frac{2i}{ca} - \zeta_{z\varphi}\,h_{ex})L \qquad (8)$$

$$\begin{aligned}V_c &= \bar{e}_\varphi\, 2\pi a n L \\ &= (-\zeta_{\varphi\varphi}\,h_{ex} + \zeta_{\varphi z}\,\frac{2i}{ca})2\pi a n L\end{aligned} \qquad (9)$$

where $L$ is the wire length and $n$ is the number of coil turns per unit length.

## III. Basic equations

The calculation of $\hat{\zeta}$ is based on the solution of the Maxwell's equations for the fields $\mathbf{e}$ and $\mathbf{h}$ together with the equation of motion for the magnetisation vector $\mathbf{M}$. An analytical treatment is possible in a linear approximation with respect to the time-variable parameters $\mathbf{e}$, $\mathbf{h}$, $\mathbf{m} = \mathbf{M} - \mathbf{M}_0$, where $\mathbf{M}_0$ is the static magnetisation. Assuming a local relationship between $\mathbf{m}$ and $\mathbf{h}$: $\mathbf{m} = \hat{\chi}\mathbf{h}$, the problem is simplified to finding the solutions of the Maxwell equations with a given ac permeability tensor

$$\hat{\mu} = 1 + 4\pi\hat{\chi}$$
$$\operatorname{curl}\mathbf{e} = j\omega(\hat{\mu}\mathbf{h})/c, \quad \operatorname{curl}\mathbf{h} = 4\pi\sigma\mathbf{e}/c \qquad (10)$$

satisfying the boundary conditions (5). Here $\sigma$ is the conductivity. Introducing the local permeability tensor $\hat{\mu}$ corresponds to neglecting exchange effects. This approximation is reasonable for not very high frequencies, such that the skin depth is still larger than the exchange length. Further assumptions about $\hat{\mu}$ are needed. The permeability depends on many factors, including the domain configuration, anisotropy and stress distribution, and the mode of magnetisation (domain wall motion or magnetisation rotation). These factors can be complex in real materials, making modelling very difficult. In this analysis, the domain structure is not considered, it can be eliminated by a proper dc bias. It is assumed that $\mathbf{M}_0$ is aligned in a helical direction having a constant angle $\theta$ with the wire axis (the details of the dc magnetic structure are given in Section VI). In this case, $\hat{\mu}$ is determined by the magnetic moment rotation and is independent of the position. This is approximation even for an ideal material, since a circumferential magnetisation near the wire centre results in an infinite exchange energy. Then, there is always a radial distribution in permeability, which is stronger in the case of a helical anisotropy due to a stress distribution. Here, when we consider a high frequency case, the permeability is predominantly a surface permeability (not effected by radial changes). In the low frequency case where the radial change in permeability becomes important the magneto-impedance effect is relatively small. In fact, an averaged value of the permeability can be used for a low frequency approximation. We extrapolate the high frequency result to the low frequency case using the same permeability parameter. The comparison between the theory and experiment is good proving that this approach is reasonable and a radial distribution in permeability is not significant for MI effects. The tensor $\hat{\mu}$ has a general form with

$$\mu_{\varphi r} = -\mu_{r\varphi}, \quad \mu_{rz} = -\mu_{zr}, \quad \mu_{\varphi z} = \mu_{z\varphi}$$

due to the magnetic symmetry. Considering that the time dependence is given by $\exp(-j\omega t)$ and utilising the cylindrical symmetry ($\mathbf{e} = (e_\varphi, e_z)$, $\mathbf{b} = (b_\varphi, b_z)$), the Maxwell's equations can be reduced to

$$\frac{\partial e_z}{\partial r} = -\frac{j\omega}{c} b_\varphi, \quad \frac{1}{r}\frac{\partial(re_\varphi)}{\partial r} = \frac{j\omega}{c} b_z \qquad (11)$$

$$\frac{\partial h_z}{\partial r} = -\frac{4\pi\sigma}{c} e_\varphi, \quad \frac{1}{r}\frac{\partial(rh_\varphi)}{\partial r} = \frac{4\pi\sigma}{c} e_z \qquad (12)$$

where $\mathbf{b} = \hat{\mu}\mathbf{h}$ is the vector of magnetic induction. Since $b_r = 0$ (which satisfies the boundary conditions at the wire surface), the material equations are of the form

$$\begin{aligned} b_\varphi &= \mu_1 h_\varphi + \mu_3 h_z \\ b_z &= \mu_3 h_\varphi + \mu_2 h_z \end{aligned} \qquad (13)$$

The magnetic parameters are given by

$$\mu_1 = \mu_{\varphi\varphi} + \mu_{\varphi r}^2 / \mu_{rr}, \quad \mu_2 = \mu_{zz} + \mu_{rz}^2 / \mu_{rr},$$

$$\mu_3 = \mu_{\varphi z} - (\mu_{\varphi r}\mu_{rz})/\mu_{rr} \qquad (14)$$

Substituting (13) to (11), (12) and eliminating the electric field $\mathbf{e}$ gives the equations for the magnetic field components $h_z$ and $h_\varphi$

$$r^2 \frac{\partial^2 h_\varphi}{\partial r^2} + r\frac{\partial h_\varphi}{\partial r} + (k_1^2 r^2 - 1)h_\varphi = -k_3^2 r^2 h_z$$

$$r^2 \frac{\partial^2 h_z}{\partial r^2} + r\frac{\partial h_z}{\partial r} + k_2^2 r^2 h_z = -k_3^2 r^2 h_\varphi \qquad (15)$$

where $k_i^2 = \mu_i (4\pi j\omega\sigma/c^2)$ and $i = 1, 2, 3$. Equations (15) are solved imposing boundary conditions (5) at the wire surface. The boundary conditions at $r = 0$ must exclude the infinite solutions, requiring

$$h_\varphi(r \le a) < \infty, \quad h_z(r \le a) < \infty \qquad (16)$$

Then, the coupled equations (15) with conditions (5) and (16) are completely determined.

In the present analysis, asymptotic solutions of equations (15) are found in two limiting cases: $\delta \ll a$ and $\delta \gg a$, where $\delta = c/\sqrt{2\pi\sigma\omega}$ is the skin depth in a non-magnetic material ($\hat{\mu} = 1$), as power series in a corresponding small parameter ($\delta/a$ or $a/\delta$). On the other hand, no condition is imposed on the value of the magnetic skin depth $\delta_i = c/\sqrt{2\pi\sigma\omega\mu_i}$, where $\mu_i$ is a corresponding magnetic parameter defined by (14). The series representation for the electric field $\mathbf{e} = (e_z, e_\varphi)$ is then deduced from equations (12). If the surface values $\bar{e}_\varphi, \bar{e}_z$ are written in the form linear with respect to the boundary values $\bar{h}_\varphi$ and $h_{ex}$, the surface impedance tensor can be calculated from equations (7).

To simplify the further analysis, it is useful to write the tensor $\hat{\mu}$ in the co-ordinate system with the axis $z' \| \mathbf{M}_0$ where it has the simplest form. In the case of a uniform precession of the total magnetisation vector $\mathbf{M}$ around $\mathbf{M}_0$, the susceptibility tensor in the prime co-ordinates $(z', \varphi', r)$ related with the equilibrium magnetisation $\mathbf{M}_0$ (see Fig. 1) is of the form

$$\hat{\chi} = \begin{pmatrix} \chi_1 & -j\chi_a & 0 \\ j\chi_a & \chi_2 & 0 \\ 0 & 0 & 0 \end{pmatrix} \qquad (17)$$

This form can be easily obtained from the linearised Landau-Lifshitz equation. The expressions for $\chi_1$, $\chi_2$, $\chi_a$ depend on a given magnetic configuration and will be determined later. The susceptibility tensor can be converted to the original co-ordinate representation $(z, \varphi, r)$ by rotating the prime system by angle $\theta$ which determines the direction of $\mathbf{M}_0$ with respect to the wire axis $z$

$$\hat{\chi} = \begin{pmatrix} \chi_1 & -j\chi_a \cos\theta & j\chi_a \sin\theta \\ j\chi_a \cos\theta & \chi_2 \cos^2\theta & -\chi_2 \sin\theta\cos\theta \\ -j\chi_a \sin\theta & -\chi_2 \sin\theta\cos\theta & \chi_2 \sin^2\theta \end{pmatrix}$$

(18)

Using (18) gives

$$\mu_1 = 1 + 4\pi\cos^2(\theta)\chi, \quad \mu_2 = 1 + 4\pi\sin^2(\theta)\chi,$$

(19)

$$m_3 = -4p\sin(q)\cos(q)\,c, \quad c = c_2 - \frac{4p\,c_a^2}{1+4p\,c_1}.$$

### III. High frequency approximation

The singular perturbation method constructed with respect to a small parameter $b = d/a \ll 1$ is used to obtain asymptotic solutions of equations (15) in the case of high frequencies. Customarily, this case is treated by considering the plane geometry. However, such approach allows the zero-order terms only to be found. For the purpose to build a general asymptotic solution valid in a wide frequency range, the higher-order terms in the series expansion are important as well.

Introducing a new variable $x = r/a$ and multiplying equations (15) by $b^2$ gives

$$b^2 x^2 \frac{\partial^2 h_\varphi}{\partial x^2} + b^2 x \frac{\partial h_\varphi}{\partial x} + \left(b_1^2 x^2 - b^2\right) h_\varphi = -b_3^2 x^2 h_z$$

$$b^2 x^2 \frac{\partial^2 h_z}{\partial x^2} + b^2 x \frac{\partial h_z}{\partial x} + b_2^2 x^2 h_z = -b_3^2 x^2 h_\varphi$$

(20)

The boundary conditions for equations (20) are

$$h_\varphi(1) = \bar{h}_\varphi, \quad h_z(1) = h_{ex}$$
$$h_\varphi(x) < \infty, \quad h_z(x) < \infty, \quad 0 \le x \le 1$$

(21)

Here $b_i^2 = 2j m_i$. Equations (20) have a small parameter at the second-order derivative and are related to so-called singular perturbed equations.[29-31] The solution of such an equation can be represented as the sum of two (regular and singular) asymptotic series of powers of the small parameter. The regular part approximates the solution within a certain internal area whereas the singular series is related to the boundary layer (near $x = 1$) where the solution undergoes rapid changes. Such a layer is named as a frontier layer. In our case it corresponds to the skin depth. In the internal area $0 < x < 1$, the singular part decays exponentially and the regular series has a smooth behaviour.

Following the singular perturbation method, the solution of (20) is written in the form

$$h_\varphi(x,\eta) = \sum_{n \ge 0} b^n R_{\varphi\,n}(x) + \sum_{n \ge 0} b^n S_{\varphi\,n}(\eta), \quad (22)$$

$$h_z(x,\eta) = \sum_{n \ge 0} b^n R_{z\,n}(x) + \sum_{n \ge 0} b^n S_{z\,n}(\eta) \quad (23)$$

where $R_{\varphi\,n}$, $R_{z\,n}$ and $S_{\varphi\,n}$, $S_{z\,n}$ represent regular and singular terms, respectively, and $\eta = (x-1)/b$ is "the fast" variable. Equations (20),(21) written in terms of the fast variable $\eta$ become

$$(\eta\beta+1)^2 \frac{\partial^2 h_\varphi}{\partial \eta^2} + \beta(\eta\beta+1)\frac{\partial h_\varphi}{\partial \eta} + \left(\beta_1^2(\eta\beta+1)^2 - \beta^2\right)h_\varphi = -\beta_3^2(\eta\beta+1)^2 h_z$$

$$(\eta\beta+1)^2 \frac{\partial^2 h_z}{\partial \eta^2} + \beta(\eta\beta+1)\frac{\partial h_z}{\partial \eta} + \beta_2^2(\eta\beta+1)^2 h_z = -\beta_3^2(\eta\beta+1)^2 h_\varphi$$

$$h_\varphi(0) = \bar{h}_\varphi, \quad h_z(0) = h_{ex}$$
$$h_\varphi(\eta) < \infty, \quad h_z(\eta) < \infty, \quad -1/\beta \le \eta \le 0$$

(24)

Substituting the regular series into (20) and the singular series into (24), and grouping together terms of the same power $n$ of $\beta$, the asymptotic solution of degree $n$ is constructed. In the case of the regular series, the zero-order ($n = 0$) approximation gives

$$b_2^2 R_{z0}(x) = -b_3^2 R_{\varphi\,0}(x)$$
$$b_1^2 R_{\varphi\,0}(x) = -b_3^2 R_{z0}(x)$$

(25)

Equations (25) are satisfied only if $R_{z0}(x) = R_{\varphi\,0}(x) = 0$. Proceeding in a similar way, it can be shown that all higher-order terms turn out to be zero as well. Therefore, in the present case the solution does not have a regular part, which could be expected as a consequence of the skin-effect. The existence of the regular solution would result in the deep "diffusion" of the electromagnetic field inside the wire at high frequencies. According to the general property of singular equations, the singular part decays exponentially as $\exp(-a(1-x)/d)$, therefore the frontier layer corresponds to the skin depth $d$.

Considering the singular series, the zero-order terms are found by solving the following equations

$$\frac{\partial^2 S_{\varphi\,0}}{\partial \eta^2} + b_1^2 S_{\varphi\,0} = -b_3^2 S_{z0}, \quad S_{\varphi\,0}(0) = \bar{h}_\varphi$$

$$\frac{\partial^2 S_{z0}}{\partial \eta^2} + b_2^2 S_{z0} = -b_3^2 S_{\varphi\,0}, \quad S_{z0}(0) = h_{ex}$$

(26)

To choose a physically reasonable solution, the following condition has to be imposed

$$\lim_{\substack{b \to 0 \\ x < 1}} S((x-1)/b) = \lim_{\eta \to -\infty} S(\eta) = 0 \quad (27)$$

The solution of (26) is taken in the form $C\,\mathbf{y}\exp(\xi\eta)$ where $\mathbf{y}$ is the intrinsic vector of coupled equations, $C$ is a constant, and $\xi$ satisfies to

$$\xi^4 + \xi^2\left(b_1^2 + b_2^2\right) + \left(b_1^2 b_2^2 - b_3^4\right) = 0 \quad (28)$$

Using $b_i^2 = 2j m_i$, where $\mu_i$ are determined by (19), we obtain

$$x_1 = \pm(1-j), \quad x_2 = \pm(1-j)\sqrt{\mu_{ef}}$$
$$\mu_{ef} = 1 + 4\pi\chi \qquad (29)$$

In (29), only sign "+" has to be taken to be consistent with condition (27) since in this case the exponent $\exp(\xi\eta)$ is limited for any $\eta < 0$. Finally, the general solution of (26) is represented as

$$\begin{pmatrix} h_\varphi \\ h_z \end{pmatrix} = C^{(1)} \begin{pmatrix} y_1^{(1)} \\ y_2^{(1)} \end{pmatrix} \exp\left(\frac{(1-j)a}{\delta}(x-1)\right)$$
$$+ C^{(2)} \begin{pmatrix} y_1^{(2)} \\ y_2^{(2)} \end{pmatrix} \exp\left(\frac{(1-j)a}{\delta}\sqrt{\mu_{ef}}(x-1)\right) \qquad (30)$$

There are two decay lengths in Eq. (30): $\delta$ and $\delta_m = \delta/\sqrt{\mu_{ef}}$. The former $\delta$ is related to a non magnetic but electrically conducting case describing the distribution of the electromagnetic field having the local polarisation with the magnetic field parallel to the dc magnetisation $\mathbf{M}_0$. The latter $\delta_m$ is a magnetic skin depth corresponding to the mode with **h** perpendicular to $\mathbf{M}_0$. In the case under consideration, the vector $\mathbf{M}_0$ is directed along the helical pass, resulting in the existence of both polarisations and the solution involving the two characteristic decay lengths.

Defining $C^{(1,2)}$ from boundary conditions in Eqs. (24), the zero-order estimate for the magnetic field $h_\varphi$, $h_z$ is completed. Substituting Eq. (30) into (12) yields the solution for the electric field **e**. Then, from Eqs. (7) the surface impedance tensor is deduced

$$\zeta = \begin{pmatrix} \zeta_{zz} & \zeta_{z\varphi} \\ \zeta_{\varphi z} & \zeta_{\varphi\varphi} \end{pmatrix}$$
$$= \zeta_0 \begin{pmatrix} \sqrt{\mu_{ef}}\cos^2(\theta) + \sin^2(\theta) & \left(\sqrt{\mu_{ef}} - 1\right)\sin(\theta)\cos(\theta) \\ \left(\sqrt{\mu_{ef}} - 1\right)\sin(\theta)\cos(\theta) & \cos^2(\theta) + \sqrt{\mu_{ef}}\sin^2(\theta) \end{pmatrix}$$
$$\zeta_0 = \frac{c(1-j)}{4\pi\sigma\delta} \qquad (31)$$

The high-frequency limit equation (31) for the surface impedance tensor (or its certain components) has been obtained in a number of papers,[15,32,33] regarding small regions at the wire surface as flat surfaces, and imposing the boundary conditions similar to (5). However, this method restricts to a zero-order approximation only. The higher-order terms can be important to determine more accurately the validity conditions. For example, it has been considered that a strong skin-effect approximation yielding (31) requires $\delta/a \ll 1$ which is much stronger than that involving the magnetic skin-depth $\delta_m/a \ll 1$. This opinion is based on the field distribution as in Eq. (30) depending on the both decay parameters. For an amorphous wire ($\sigma = 10^{16}$ s$^{-1}$) of 30 μm diameter the non-magnetic skin depth becomes of the order of $a$ at GHz frequencies. On the other hand, numerous experimental results on MI are concerned with frequencies of 1–100 MHz, and it seems that the high frequency case has a very limited use. Within the proposed method, the full asymptotic series can be found. Considering the first-order approximation is important in context to prove that the condition $\delta_m/a \ll 1$ is sufficient to justify the use of Eq. (31).

The first-order equations for $S_{z1}$ and $S_{\varphi 1}$ are of the form

$$\frac{\partial^2 S_{z1}}{\partial \eta^2} + \beta_2^2 S_{z1} = -\beta_3^2 S_{\varphi 1} - \frac{\partial S_{z0}}{\partial \eta}, \qquad S_{z1}(0) = 0$$
$$\frac{\partial^2 S_{\varphi 1}}{\partial \eta^2} + \beta_1^2 S_{\varphi 1} = -\beta_3^2 S_{z1} - \frac{\partial S_{\varphi 0}}{\partial \eta}, \qquad S_{\varphi 1}(0) = 0$$
$$(32)$$

Since the functions $\partial S_{z0}/\partial \eta$ and $\partial S_{\varphi 0}/\partial \eta$ are represented in the exponential form, the particular solution of equation (32) is given by

$$\tilde{S}_{z1} = (a_1\eta + b_1)e^{x_1\eta} + (a_2\eta + b_2)e^{x_2\eta}$$
$$\tilde{S}_{\varphi 1} = (c_1\eta + d_1)e^{x_1\eta} + (c_2\eta + d_2)e^{x_2\eta} \qquad (33)$$

where $x_{1,2}$ are determined by Eqs. (29). The general solution of coupled homogeneous equations (32) is of the form of (30) where the constants $C^{(1)}$ and $C^{(2)}$ are found from the zero boundary conditions in Eqs. (32). The calculation process is straight forward but time consuming and results in rather cumbersome expressions. However, substituting the values of $\beta_i$ specific for the given problem, the result becomes as simple as

$$\left.\frac{\partial S_{z1}}{\partial \eta}\right|_{\eta=0} = -\frac{1}{2}h_z, \qquad \left.\frac{\partial S_{\varphi 1}}{\partial \eta}\right|_{\eta=0} = \frac{1}{2}h_\varphi \qquad (34)$$

Then, the first-order term for the impedance tensor is

$$\zeta_1 = \frac{c(1-j)}{4\pi\sigma\delta}\left(\frac{\delta}{a}\right)\begin{pmatrix} \frac{(1+j)}{4} & 0 \\ 0 & -\frac{(1+j)}{4} \end{pmatrix} \qquad (35)$$

Comparing (35) and (31), it is seen that the ratio of $\zeta_1/\zeta_0$ is of the order $(\delta/a)/\sqrt{\mu_{ef}}$ or $\delta_m/a$. Therefore, the actual parameter in the expansion for the impedance is $\delta_m/a$, proving the validity of the high frequency results in a wider frequency region if $m_{ef}$ is sufficiently large.

## IV. Low frequency approximation

Let us now construct the solution for the impedance in the opposite limit $a/d \ll 1$. Having high-frequency result (35), it can be expected that in this case the actual parameter of the expansion involves the magnetic skin depth as well. Then, it may be difficult to join the two asymptotes together. Therefore, we would like to build the low frequency asymptote such that it could be expanded to the case $a/d_m > 1$. The solution of (15) is taken in the form

$$h_j = \bar{h}_j \frac{J_1(k_1 a x)}{J_1(k_1 a)} + \tilde{h}_j(x),$$
$$h_z = h_{ex} \frac{J_0(k_2 a x)}{J_0(k_2 a)} + \tilde{h}_z(x) \qquad (36)$$

where $J_{0,1}$ are the Bessel functions of the first kind. In (36), the first terms give the exact solutions for the homogeneous forms of equations (15). This representation for fields $h_j$, $h_z$ is proving to be adequate to get almost a monotonic transition from one asymptote to the other, changing a frequency or an external magnetic field. The functions $\tilde{h}_j$ and $\tilde{h}_z$ determining the extent of coupling of equations (5) are found from

$$x^2 \frac{\partial^2 \tilde{h}_j}{\partial x^2} + x \frac{\partial \tilde{h}_j}{\partial x} + (b_1^2 b^2 x^2 - 1)\tilde{h}_j$$
$$= -h_{ex} \frac{b_3^2 b^2 x^2 J_0(b_2 b x)}{J_0(b_2 b)} - b_3^2 b^2 x^2 \tilde{h}_z$$
$$x^2 \frac{\partial^2 \tilde{h}_z}{\partial x^2} + x \frac{\partial \tilde{h}_z}{\partial x} + b_2^2 b^2 x^2 \tilde{h}_z \qquad (37)$$
$$= -\bar{h}_j \frac{b_3^2 b^2 x^2 J_1(b_1 b x)}{J_1(b_1 b)} - b_3^2 b^2 x^2 \tilde{h}_j$$

satisfying the conditions

$$\tilde{h}_j(1) = 0 \quad \tilde{h}_z(1) = 0$$
$$\tilde{h}_j(x) < \infty \quad \tilde{h}_z(x) < \infty$$

Here we use the same notation $b = a/d$ for the small parameter though it is inverse to that used in Section IV. The solution of equation (37) is represented in terms of the asymptotic series of powers of $\beta$, using the regular perturbation method

$$\tilde{h}_j(x) = \sum_{n \geq 0} \beta^n \tilde{h}_{j\,n}(x),$$
$$\tilde{h}_z(x) = \sum_{n \geq 0} \beta^n \tilde{h}_{z\,n}(x),$$
$$\frac{J_1(b_1 b x)}{J_1(b_1 b)} = x\left[1 + \frac{b_1^2 b^2}{8}(1 - x^2)\right] + O(b^4),$$

$$\frac{J_0(b_2 b x)}{J_0(b_2 b)} = \left[1 + \frac{b_2^2 b^2}{4}(1 - x^2)\right] + O(b^4) \qquad (38)$$

Substituting series (38) into equation (37) and grouping together terms having the same powers of $\beta$, the equations for the regular series $\tilde{h}_{j\,n}(x)$ and $\tilde{h}_{z\,n}(x)$ are obtained. It turns out that the terms of the zero degree and of any (2n+1)-th degree are equal to zero. The second-degree terms are found from

$$x^2 \frac{\partial^2 \tilde{h}_{j\,2}}{\partial x^2} + x \frac{\partial \tilde{h}_{j\,2}}{\partial x} - \tilde{h}_{j\,2} = -h_{ex} b_3^2 x^2$$
$$x^2 \frac{\partial^2 \tilde{h}_{z2}}{\partial x^2} + x \frac{\partial \tilde{h}_{z2}}{\partial x} = -\bar{h}_j b_3^2 x^3 \qquad (39)$$
$$\tilde{h}_{j\,2}(1) = 0 \quad \tilde{h}_{z2}(1) = 0$$
$$\tilde{h}_{j\,2}(x) < \infty \quad \tilde{h}_{z2}(x) < \infty$$

Solving equations (39) gives
$$\tilde{h}_{j\,2} = \frac{h_{ex} b_3^2 (x - x^2)}{3}, \quad \tilde{h}_{z2} = \frac{\bar{h}_j b_3^2 (1 - x^3)}{9} \qquad (40)$$

The equations for the fourth-degree terms are

$$x^2 \frac{\partial^2 \tilde{h}_{j\,4}}{\partial x^2} + x \frac{\partial \tilde{h}_{j\,4}}{\partial x} - \tilde{h}_{j\,4}$$
$$= -h_{ex} \frac{b_2^2 b_3^2}{4} x^2(1 - x^2) - x^2(b_3^2 \tilde{h}_{z\,2} + b_1^2 \tilde{h}_{j\,2})$$
$$x^2 \frac{\partial^2 \tilde{h}_{z\,4}}{\partial x^2} + x \frac{\partial \tilde{h}_{z\,4}}{\partial x}$$
$$= -\bar{h}_j \frac{b_1^2 b_3^2}{8} x^3(1 - x^2) - x^2(b_2^2 \tilde{h}_{z\,2} + b_3^2 \tilde{h}_{j\,2})$$
$$\tilde{h}_{j\,4}(1) = 0 \quad \tilde{h}_{z\,4}(1) = 0$$
$$\tilde{h}_{j\,4}(x) < \infty \quad \tilde{h}_{z\,4}(x) < \infty$$
$$\qquad (41)$$

Substituting (40) into (41) gives

$$\tilde{h}_{j\,4} = -\bar{h}_j \frac{b_3^4}{9}\left(\frac{x^2}{3} - \frac{x^5}{24} - \frac{21}{72}x\right) -$$

$$h_{ex}\frac{b_2^2 b_3^2}{4}\left(\frac{x^2}{3} - \frac{x^4}{15} - \frac{12}{45}x\right) - h_{ex}\frac{b_1^2 b_3^2}{3}\left(\frac{x^3}{8} - \frac{x^4}{15} - \frac{7}{120}x\right)$$

$$\tilde{h}_{z4} = -\bar{h}_j \frac{b_1^2 b_3^2}{8}\left(\frac{x^3}{9} - \frac{x^5}{25} - \frac{16}{225}\right) -$$

$$\bar{h}_j \frac{b_3^2 b_2^2}{9}\left(\frac{x^2}{4} - \frac{x^5}{25} - \frac{21}{100}\right) - h_{ex}\frac{b_3^4}{3}\left(\frac{x^3}{9} - \frac{x^4}{16} - \frac{7}{144}\right)$$

(42)

equations (36),(40) and (42) give the asymptotic series for the magnetic field, as far as the $b^4$-terms

$$h_j = \bar{h}_j \frac{J_1(k_1 a\, x)}{J_1(k_1 a)} + b^2 \tilde{h}_{j\,2}(x) + b^4 \tilde{h}_{j\,4}(x)$$

$$h_z = h_{ex}\frac{J_0(k_2 a\, x)}{J_0(k_2 a)} + b^2 \tilde{h}_{z\,2}(x) + b^4 \tilde{h}_{z\,4}(x)\quad (43)$$

Calculating the electric field from Eqs. (12) and representing it in the form linear in $\bar{h}_j$ and $h_{ex}$, the components of the surface impedance tensor are obtained

$$\zeta_{zz} = \frac{k_1 c}{4\pi\sigma}\frac{J_0(k_1 a)}{J_1(k_1 a)} + \frac{1}{54}\left(\frac{a}{d}\right)^4 \frac{c\,\mu_3^2}{\pi\sigma a} \quad (44)$$

$$\zeta_{\varphi\varphi} = -\frac{k_2 c}{4\pi\sigma}\frac{J_1(k_2 a)}{J_0(k_2 a)} + \frac{1}{36}\left(\frac{a}{d}\right)^4 \frac{c\,\mu_3^2}{\pi\sigma a} \quad (45)$$

$$\zeta_{\varphi z} = \zeta_{z\varphi} = j\frac{a\omega}{3c}\mu_3 - \left(\frac{a}{d}\right)^4\left[\frac{\mu_1\mu_3}{60} + \frac{\mu_2\mu_3}{30}\right]\frac{c}{\pi\sigma a}$$

(46)

The second terms in Eqs. (44)–(46) depend on the corresponding magnetic parameters $\mu_i$, demonstrating that the actual expansion parameter involves a sort of magnetic skin depth (but not exactly $\delta_m$). For example, in equation (44) in the case of $k_1 a >> 1$ (but $a/\delta << 1$) the ratio of the second term to the first becomes $(1/54)(k_3 a)^4 / k_1 a$. The values of $k_i$ are of the same order, as it follows from Eqs. (15),(19). Yet, the numerical analysis shows that the first terms in Eqs. (44)–(46) can give the main contribution to the impedance even in the case of $(k_i / a) \approx 1$, which is illustrated by a small numerical factor $1/54$ in the above example. This helps joining the low frequency asymptote with the high frequency one. In the next Section, the asymptotic behaviour will be discussed in more detail for different magnetic configurations.

### V. Analysis of the impedance behaviour for two types of anisotropy

Our approach can be applied to a wire having a circumferential or helical anisotropy. In general, the anisotropy axis $\mathbf{n}_K$ has an angle $45° < \alpha \leq 90°$ with the wire axis (z-axis), as shown in Fig. 1. The wire is assumed to be in a single domain state with the static magnetisation $\mathbf{M_0}$ directed in a helical way having an angle $\theta$ with the z-axis. The radial variation in $\theta$ is neglected as explained in Section III. The magnetic configuration changes under the application of the external axial magnetic field $H_{ex}$ and the dc bias current $I_b$ inducing the circular magnetic field $H_b$. The stable direction of $\mathbf{M_0}$ is found by minimising the energy $U$

$$\partial U / \partial \theta = 0$$

$$U = -K\cos^2(\alpha - \theta) - M_0 H_{ex}\cos\theta - M_0 H_b \sin\theta$$

(47)

where $K$ is the anisotropy constant, $H_b$ is the dc circular field induced by the current $I_b$. Equation (47) describes Fig. 3. Magnetisation curves $M_{0z}(H_{ex})$ for different magnitudes of the dc bias field $H_b$. The cases related to a circumferential ($\alpha = 90°$) and helical ($\alpha = 50°$) anisotropy are shown in (a) and (b), respectively the rotational magnetisation process demonstrated in Fig. 3 where the magnetisation plots for two types of anisotropy and different values of the dc bias $H_b$ are given. The domain processes may not be essential for the reversal of $\mathbf{M_0}$, since the magnetisation vector during its rotation is held parallel to the surface, without going through high-energy demagnetisation states. In the case of a circumferential anisotropy (Fig. 3(a)), a dc bias current (inducing $H_b$ larger than the coercivity) eliminates the domain structure, without changing the magnetic symmetry. The case of a helical anisotropy is more complicated. The dc bias causes the transition from a symmetric hysteresis curve to asymmetric anhysteretic one which happens at $H_b/H_K = \cos\alpha$ (see Fig. 3(b)). Therefore, in this case a much larger bias field is needed to realise a single domain state.

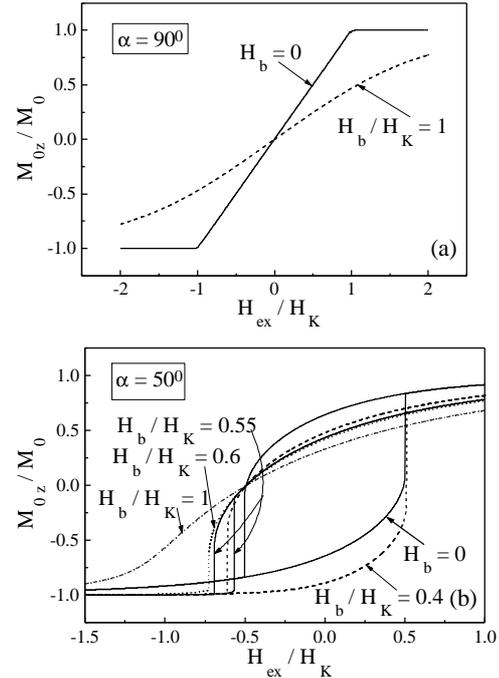

The permeability tensor $\hat{\mu} = 1 + 4\pi\hat{\chi}$ is found from a linearised Landau-Lifshitz equation for $\mathbf{m} = \hat{\chi}\mathbf{h}$ written in the co-ordinate system $(z', \varphi', r)$ with the axis $z'$ parallel to $\mathbf{M}_0$

$$-j\omega\mathbf{m} + (\omega_H - j\tau\omega)(\mathbf{m}\times\mathbf{n}_{z'}) + \gamma M_0((\hat{N}\mathbf{m})\times\mathbf{n}_{z'})$$
$$= \gamma M_0(\mathbf{h}\times\mathbf{n}_{z'})$$

(48)

Where $\omega_H = \gamma(\partial U/\partial \mathbf{M}_0)_{z'}$, $\gamma$ is the gyromagnetic constant, $\tau$ is the spin-relaxation parameter, $\hat{N}$ is the tensor of the effective anisotropy factors in $(z', \varphi', r)$-system

$$N_{z'z'} = -\frac{2K}{M_0^2}\cos^2(\theta - \alpha),$$
$$N_{\varphi'\varphi'} = -\frac{2K}{M_0^2}\sin^2(\theta - \alpha),$$
$$N_{z'\varphi'} = N_{\varphi'z'} = \frac{K}{M_0^2}\sin 2(\theta - \alpha) \quad (49)$$

Solving equation (48) determines the susceptibility tensor $\hat{\chi}$ which has the form of Eq. (17) with

$$\chi_1 = \omega_M(\omega_1 - j\tau\omega)/D,$$
$$\chi_2 = \omega_M(\omega_2 - j\tau\omega)/D, \quad \chi_a = \omega\omega_M/D,$$
$$D = (\omega_2 - j\tau\omega)(\omega_1 - j\tau\omega) - \omega^2,$$
$$\omega_1 = \gamma[H_{ex}\cos\theta + H_b\sin\theta + H_K\cos 2(\alpha - \theta)],$$
$$\omega_2 = \gamma[H_{ex}\cos\theta + H_b\sin\theta + H_K\cos^2(\alpha - \theta)],$$
$$H_K = 2K/M_0 \quad \omega_M = \gamma M_0.$$

(50)

The impedance tensor is determined via the permeability parameters $\mu_i$ (low-frequency case) or the parameter $\mu_{ef}$ (high-frequency case), all of them are determined by the apparent susceptibility $\chi$ in Eqs. (19). Substituting (50) into (19) gives

$$\chi = \frac{\omega_M(\omega_2 - j\tau\omega) + 4\pi\omega_M^2}{(\omega_1 - j\tau\omega)(\omega_2 + 4\pi\omega_M - j\tau\omega) - \omega^2}$$

(51)

Equation (51) shows that the resonance change in $\chi$ can be expected at rather high frequencies (the resonance frequency is roughly equal to $\gamma\sqrt{H_K 4\pi M_0}/2\pi \sim 500$ MHz for $H_K = 5$ Oe,

$4\pi M_0 = 6000$ G). Then, the MI effects at frequencies of 1–100 MHz are not related to the ferromagnetic resonance. This statement is important since in a number of recent works [34-36] MI characteristics are explained exclusively as a consequence of the resonance behaviour of the permeability. Yet, a high sensitivity of $\chi$ with respect to $H_{ex}$ is needed to obtain large impedance changes. This can be realised by changing the direction of $\mathbf{M}_0$ under the effect of the field. As follows from Fig. 3, the magnetisation angle changes for fields of the order of the anisotropy field $H_K$, which is also the region of the major change in the permeability and the impedance. For higher fields, $\chi$ changes little resulting in insensitive impedance behaviour. Therefore, the overall reason of the MI effects is the redistribution of the high-frequency current density when the static magnetic structure is changed.

Having specified the static magnetic configuration and the ac permeability tensor, we can proceed with the impedance analysis, using equations (44)–(46) for the low frequency case or equations

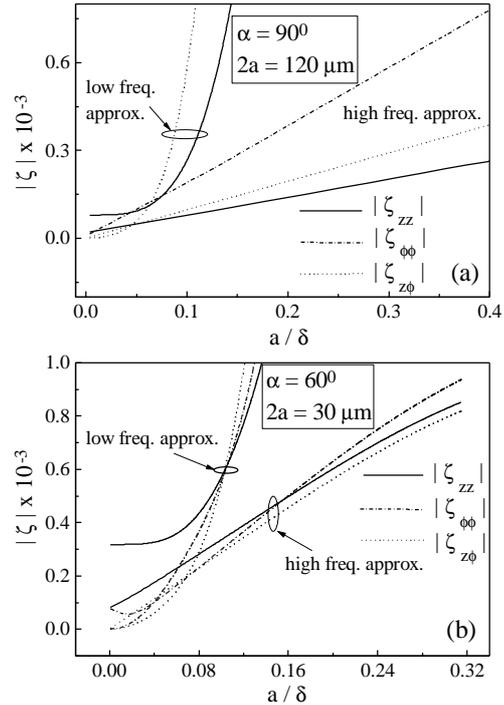

Fig. 4. Frequency spectra of the components of tensor $\hat{\zeta}$ calculated using the low and high frequency approximations for $\alpha = 90°$ in (a) and $\alpha = 60°$ in (b). $H_{ex} = 0.25 H_K$, $H_b = 0$. Parameters used: $H_K = 5$ Oe, $\sigma = 10^{16}$ sec$^{-1}$, $M_0 = 500$ G, $\tau = 0.2$, $\gamma = 2\cdot 10^7$ rad/s Oe.

(31),(35) for the opposite limit. Since both the approximations involve as an actual expansion parameter a certain magnetic skin depth, the choice between them depends not only on the value of frequency, but also on the value of $H_{ex}$ determining the permeability parameters. Figure 4 shows the components of the impedance tensor as functions of the expansion parameter $a/\delta$ (or as functions of frequency) for $H_{ex} = 0.25 H_K$ and two anisotropies: circumferential ($\alpha = 90°$) and helical ($\alpha = 60°$). For these parameters, the values of the permeability are fairly large and the transition from the low-frequency case to the high-frequency one occurs at $a/\delta = 0.04 - 0.08$. For $V_{zj}$, $V_{jj}$ components, the two asymptotes have an intersection regions (or even for $\alpha = 90°$, $V_{jj}$ monotonically transits to the high frequency case), for $V_{zz}$ there is a certain gap, actually rather small, but a sort of interpolation is needed. Considering the field dependencies of the impedance tensor, a practical rule to replace a low frequency asymptote by the high frequency one may be the condition that the second term in expansions (44)−(46) has grown up to 10% of the first one.

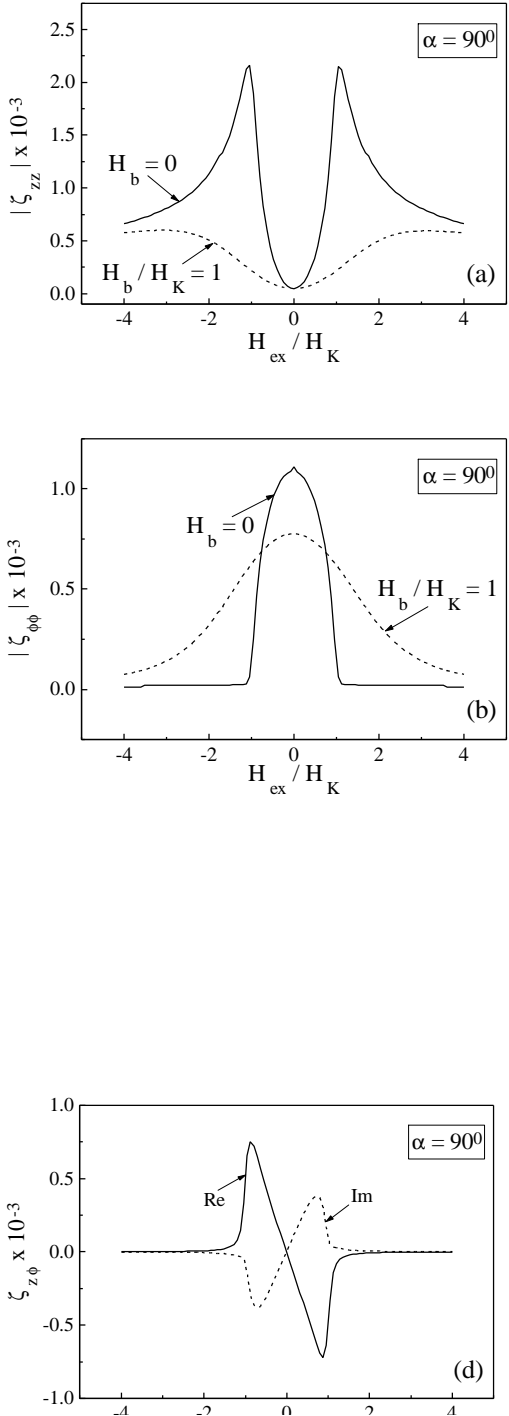

(a), (b), (d) panels

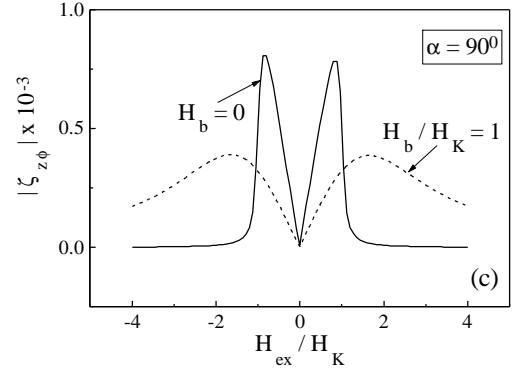

(c) panel

Fig. 5. Field characteristics of the components of tensor $V$ for a circumferential anisotropy. In (a)-(c) plots of magnitude of $V_{zz}$, $V_{jj}$ and $V_{zj}$ vs. $H_{ex}$, respectively, are given for $H_b/H_K = 0$ and 1. In (d), real and imaginary parts of $V_{zj}$ vs. $H_{ex}$ are plotted for $H_b = 0$. $2a = 120$ μm, $f = 20$ MHz.

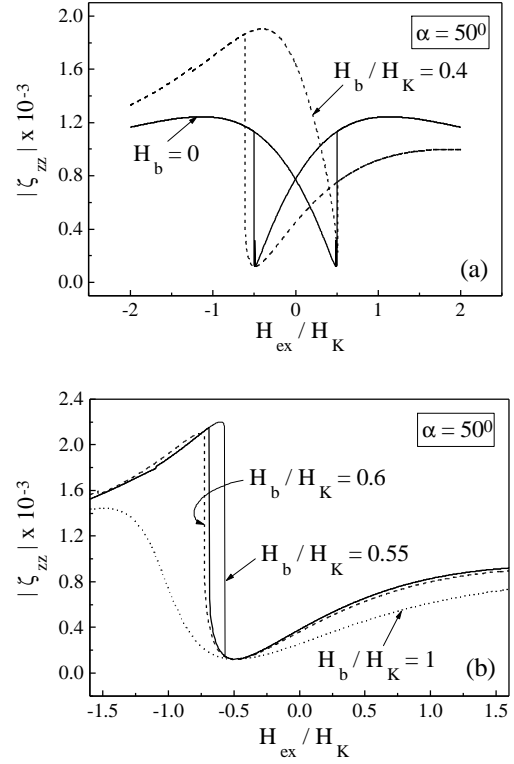

Fig. 6. Modifications of the longitudinal impedance $V_{zz}$ vs. $H_{ex}$ under the effect of the dc bias $0 \leq H_b/H_K \leq 1$. $\alpha = 50°$.

The field characteristics of the impedance tensor are determined by the combined effect of $c(H_{ex})$ and $q(H_{ex})$, and are presented in Figs. 5–8 for the two types of anisotropy. The case of the circumferential anisotropy ($\alpha = 90°$) is given in Fig. 5. For this case, $H_{ex}$ is a hard axis field, then both $M_{0z}(H_{ex})$ and $c(H_{ex})$ do not exhibit a hysteresis. The positions of maximums for $V_{zz}$, $V_{\varphi\varphi}$, $V_{z\varphi}$ ($=\varsigma_{\varphi z}$) are closely related to those for $\cos^2 q$, $\sin^2 q$, $\sin 2q$, namely, $|H_{ex}|=H_K$, $0$, $H_K/2$, respectively. With increasing frequency, the peaks for $V_{zz}$ and $V_{z\varphi}$ shift towards higher fields which is related to the permeability spectra. The application of the circular bias $H_b$ makes the peaks smaller and broader, but does not lead to a characteristically different behaviour. The diagonal components $V_{zz}$ and $V_{\varphi\varphi}$ are symmetrical with respect to $H_{ex}$, whereas the off-diagonal components $V_{z\varphi}$ or $V_{\varphi z}$ are anti-symmetrical, which is demonstrated in Fig. 5(d) by plotting the real and imaginary parts of $V_{z\varphi}$.

The case of a helical anisotropy ($a = 50°$) is more complicated involving hysteresis and considerable modifications under the effect of $H_b$. Analysing the behaviour of $V_{zz}$ vs. $H_{ex}$, shown in Fig. 6, we see that as the field decreases from positive value, $V_{zz}$ exhibits a broad flat peak which occurs between 0 and $H_K$, depending on the anisotropy angle $a$. Upon reversing the field direction, the impedance rapidly drops down to its original low value, exhibiting the highest sensitivity. With further increase in $H_{ex}$, it jumps back to the level seen for positive fields, which is associated with irreversible rotational flip in $\mathbf{M}_0$. With increasing the dc bias $H_b$, considerable asymmetry appears in the impedance plots. Further increase in $H_b$ results in a sudden shift of the hysteresis to negative fields with a simultaneous shrinkage of the hysteresis area, and $H_b > H_K \cos a$ results in the disappearance of the hysteresis. For $H_b$ slightly larger than $H_K \cos a$, the field sensitivity of the impedance change is especially high: for negative fields the nominal change can be more than 100% when $H_{ex}$ is changed by only $0.1 H_K$. The other components of the impedance tensor show characteristically similar behaviour under the effect of $H_b$, as demonstrated in Fig. 7 and Fig. 8. Similar results have been obtained for the case of MI in crossed-anisotropy multilayers.[28,37]

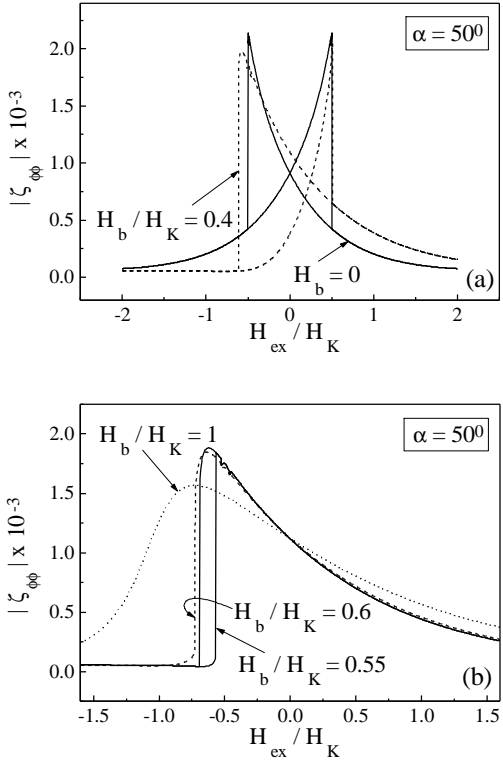

Fig. 7. Modifications of the circumferential impedance $V_{\varphi\varphi}$ vs. $H_{ex}$ under the effect of the dc bias $0 \leq H_b/H_K \leq 1$. $a = 50°$.

### VII. Experimental results and comparison with the theory

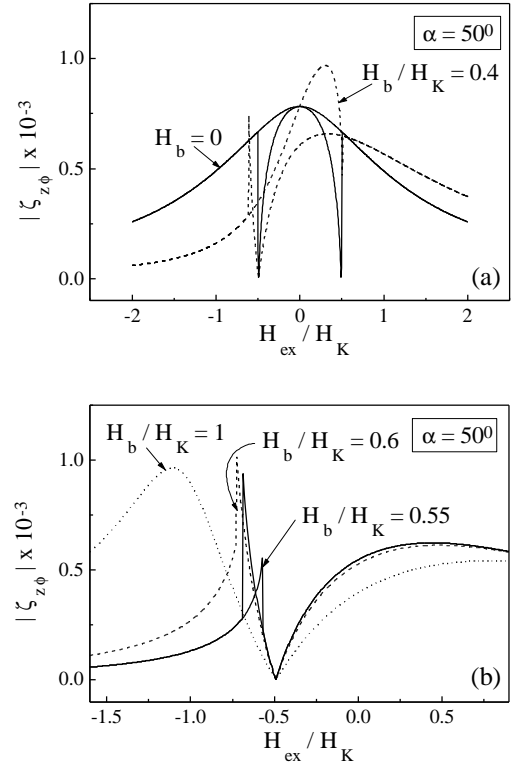

Fig.8. Modifications of the off-diagonal impedance $V_{z\varphi}$ vs. $H_{ex}$ under the effect of the dc bias $0 \leq H_b/H_K \leq 1$. $a = 50°$.

An important next step is to compare the theoretical impedance characteristics with those obtained experimentally. The experimental research on magneto-impedance in amorphous wire, although rather wide is mainly restricted to measurements of the voltage across the wire, which corresponds to measuring $V_{zz}$. A number of results reported by different groups on $V_{zz}(H_{ex})$ seem to be in conflict.

The field characteristics for same frequencies, obtained for similar wire samples, can exhibit completely different behaviour. This is a consequence of different ac excitations used, resulting in different magnetisation mechanisms involved in each case. For example, in the case of a circular or a helical domain structure, the ac current may cause the irreversible domain movement if its amplitude is larger than that corresponding to the circular coercivity. Such irreversible domain processes take place even at frequencies of few MHz. This process will mainly determine the field behaviour of the impedance: $V_{zz}(H_{ex})$ has a maximum at $H_{ex} = 0$ and decreases rapidly with increasing the field.[6-8] This is due to the corresponding behaviour of the differential domain permeability under the effect of a hard axis field. If the current amplitude is small and irreversible domain displacements are not possible, the longitudinal impedance has two symmetrical maximums at $H_{ex} \approx H_K$, in agreement with that shown in Fig. 5.[37,38] Regarding the other components of the impedance tensor there are just few experimental works on field characteristics of $V_{z\varphi}$ and $V_{\varphi\varphi}$ for a wire with circumferential anisotropy.[38,39]

For the sake of accurate quantitative comparison, we have carried out measurements of the full surface impedance tensor as a function of $H_{ex}$, at conditions as close corresponding to the theoretical model as possible. Care has been taken to realise a linear ac excitation (the amplitudes of ac currents, magnetisation and fields are considerably smaller than such dc parameters as the coercivity, anisotropy, dc magnetisation). Another model restriction is considering a single domain state. In the experiment, the domain structure can be eliminated by a dc current, however, in the case of a helical anisotropy, the field produced by this current has to be larger than the anisotropy field (not coercivity). In the cases, where domain structure is inevitable, the effect of domain wall dynamics on impedance behaviour is less at higher frequencies due to damped wall motion.

Two kinds of wires have been used: as-cast 120 $\mu m$ diameter CoFeSiB wire having a nearly zero magnetostrictive constant and a circumferential anisotropy (at least in the outer region), and tension-annealed 30 $\mu m$ diameter CoSiB wire (magnetostriction $\lambda = -3 \cdot 10^{-6}$) having a spontaneous helical anisotropy due to a residual stress distribution.[40,41]

**A. Experimental method**

The surface impedance tensor $V$ is found via measuring $S_{21}$ parameter by a Hewlett–Packard (4195A) 2–channel Network/Spectrum, which is represented by the ratio of the forward transmission signal $V_T$ to the excitation signal $V_S$. The frequency of the ac source is fixed, and the voltage of the dc source is used as a sweep parameter. The signal from the dc source is amplified with the dc power amplifier that supplies power to the coil inducing the external field $H_{ex}$. The longitudinal diagonal component $V_{zz}$ is determined by the usual way, measuring the wire voltage $V_w$ when it is excited by the ac wire current (Fig. 2(a) with $h_{ex} = 0$). In this case, in equation (8) $h_{ex} = 0$, with the result that $S_{21} = V_w/V_S = V_{zz}(H_{ex})(\bar{h}_\varphi L/V_S)$. The circumferential diagonal component $V_{\varphi\varphi}$ corresponds to the voltage $V_c$ in the secondary coil mounted on the wire which is excited by the ac axial magnetic field induced in the primary coil (Fig. 2(b) with $i = 0$). In this case,

$$V_c = j\omega n_2 L h_{ex} p(a_2^2 - a^2)/c - 2\pi a n_2 L h_{ex} V_{\varphi\varphi},$$
(52)

Here $a_2$ is the radius of the secondary coil and $n_2$ is a number of its turns per unit length. In equation (52), the first term represents the contribution from the flux between the wire and the secondary coil (the flux through the air gap), the second term corresponds to the coil voltage defined by equation (9) with $\bar{h}_\varphi = 0$. For wires having sufficiently large diameter (few tens of microns) it is quite possible to wind the secondary coil directly on the wire. In this case, the flux through the air gap is nearly zero and there is no a large disturbance constant signal. The off-diagonal components $V_{z\varphi}$ and $V_{\varphi z}$ can be determined by measuring the coil voltage $V_c$ when the wire is excited by the ac current, or measuring the wire voltage $V_w$ in the presence of the ac axial magnetic field. The latter is used here (Fig. 2(a) with $i = 0$). In this case, in equation (8) $h_\varphi = 0$ with the result that $S_{21} = V_w/V_S = -V_{z\varphi}(H_{ex})(h_{ex} L/V_S)$.

The coil length in all the experiments is about 3 mm and the wire length is about 6 mm. The secondary coil is mounted directly on the wire: $a_2 = a$. The primary coil is mounted on a glass tube with a diameter of 1 mm. The number of turns in both coils is 30. The amplitudes of the ac excitation current (in the wire or in the coil) are chosen to be less then 1 mA, then, the non-linear ac magnetisation processes like irreversible domain displacements are not possible. The experimental studies are made with the effect of the dc current which effectively governs the static magnetic structure, as discussed.

**B. Circumferential anisotropy**

First we consider the impedance characteristics in a wire with a circumferential anisotropy $\alpha = 90^0$ and a circular domain structure in the outer region. Some of these results have been reported in Ref. 38. The experimental field dependencies for the $V_{zz}$ component and the comparison with the model calculations are shown in Figs. 9. The normalised impedance corresponds to the ratio $V_w/V_S$. The real and imaginary parts of this ratio are given in Fig. 9(a), showing two symmetrical peaks at $H_{ex}$ nearly equal to the anisotropy field $H_K \approx 5$ Oe (the value of the anisotropy field has been checked by measuring the dc magnetisation loops). When the dc bias is applied, the impedance value at zero field becomes considerably smaller. The dc current eliminates the domain structure, resulting in the decrease in the overall permeability. For not very high $I_b$, the values of the impedance at the maximuma are almost constant since they are determined by the rotational processes only. However, if $I_b$ is further increased the value of the impedance at the maximuma becomes considerably smaller and the sensitivity drops, resulting from an increase in magnetic hardness by $I_b$. Figs. 9(b),(c) give the comparison of the experimental and theoretical results. The two curves are matched at positive (or negative) saturation, therefore the theoretical values are given in $S_{21}$-units. For $I_b = 0$, the main discrepancy between the theory and experiment is for fields $H_{ex}$ smaller than the anisotropy field $H_K$, which is due to the contribution of the domain wall dynamics (which is essential even for frequency of 20 MHz) to the total permeability. The theoretical model considering a single-domain state ignores the domain dynamics completely. Applying a sufficiently large current $I_b = 100$ mA eliminates domains, and the theoretical curve becomes closer to the experimental one. Figure 10 presents the longitudinal impedance for higher frequency of 100 MHz, showing a much better agreement between the experiment and theory, since the domain walls are stronger damped and give considerably smaller contribution to the total permeability.

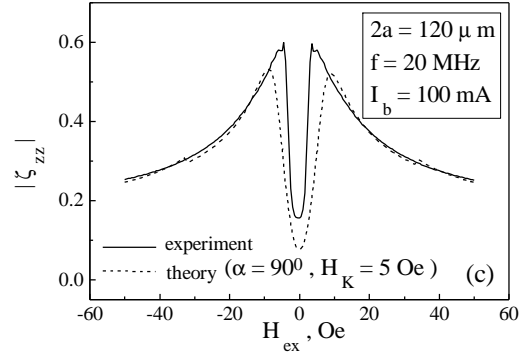

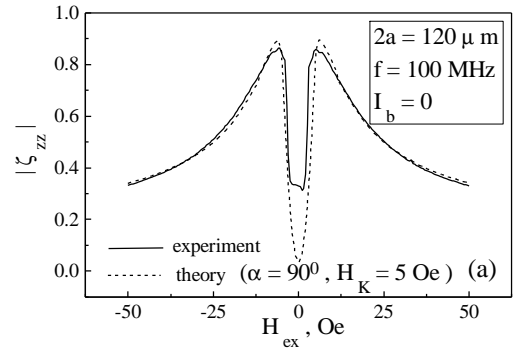

Fig. 9 Experimental plots of the longitudinal impedance $V_{zz}(H_{ex})$ for different values of $I_b$ and comparison with the theory. In (a), real and imaginary parts of the voltage ratio $V_w/V_S$ (which is proportional to $V_{zz}$) are given. In (b) and (c) the impedance magnitude $|V_{zz}|$ vs. $H_{ex}$ (in values of $|V_w/V_S|$) is compared with the theoretical dependence for a frequency of 20 MHz.

Figures 11 is related to the analysis of the circumferential diagonal impedance $\varsigma_{\varphi\varphi}$. Figure 11(a) presents the normalised voltage $V_c/V_S$ in the secondary coil mounted directly on the wire which is

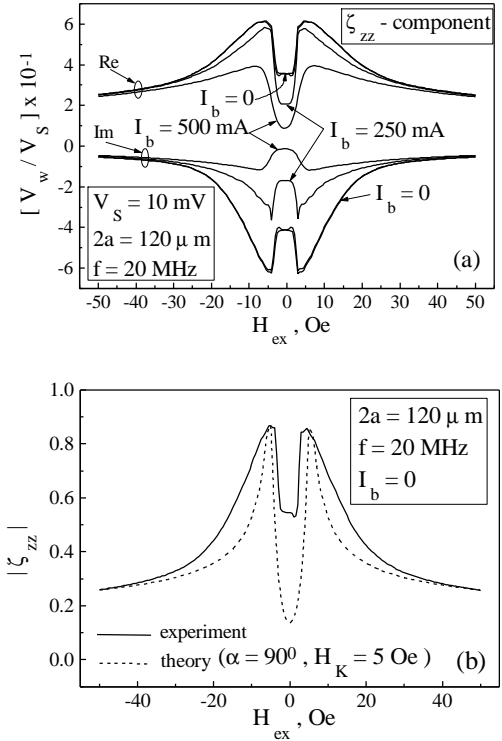

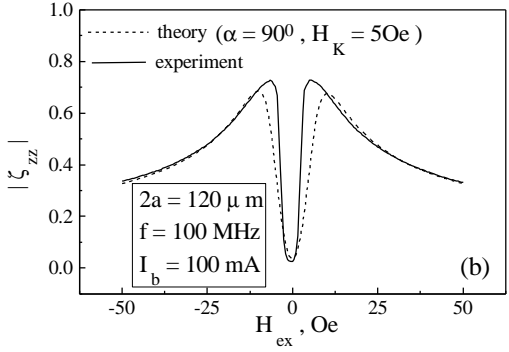

Fig.10. Theoretical and experimental plots of $|V_{zz}|$ vs. $H_{ex}$ (in values of $|V_w/V_S|$) for a frequency of 100 MHz, for $I_b = 0$ in (a) and $I_b = 100$ mA in (b).

excited by the ac axial magnetic field induced in the primary coil. This ratio is proportional to $V_{jj}$ which has a maximum at zero field and it decreases rapidly near the anisotropy field $H_K \approx 5$ Oe, whereas there is an insensitive wide region between $\pm H_K$, which is more pronounced for $I_b = 0$. It seems that this insensitive area is determined by the demagnetising factor since the sample has a rather small length (6 mm) comparing to the diameter (120 μm). However, we could not see this behaviour considering the field plots of $V_{zz}$. More probably, it is related to the combined effect of the rotational permeability (which has a maximum at zero field and is decreasing with the field) and the domain wall permeability (which has a minimum at zero field and is increasing with the field). The theoretical curve does not have this flat portion, as shown in Fig. 11(b). The application of a relatively small current $I_b = 5.57$ mA increases the sensitivity of the impedance characteristics, which may be due to a better defined circumferential magnetisation induced by this current when $\theta$ is equal almost exactly to $90^0$

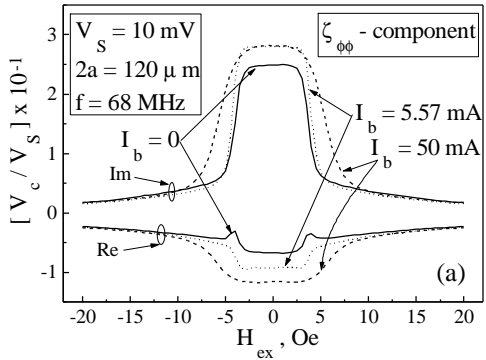

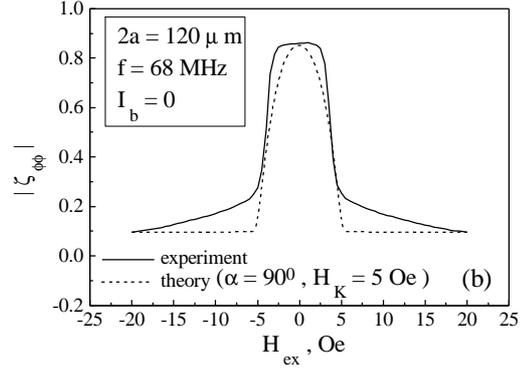

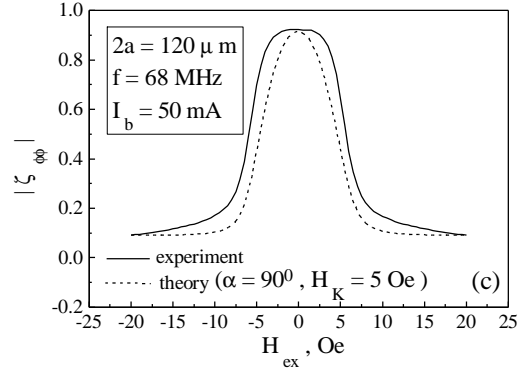

Fig. 11. Experimental plots of the circumferential impedance $V_{jj}(H_{ex})$ for different values of $I_b$ and comparison with the theory. In (a), real and imaginary parts of the voltage ratio $V_c/V_S$ (which is proportional to $V_{jj}$) are given. In (b) and (c) the impedance magnitude $|\zeta_{\varphi\varphi}|$ vs. $H_{ex}$ (in values of $|V_c/V_S|$) is compared with the theoretical dependence for a frequency of 68 MHz

without the anisotropy dispersion. The insensitive region becomes smaller under the effect of a larger $I_b$ as the domain contribution is less essential, and this case is in a good agreement with the theoretical plot as demonstrated in Fig. 11(c). Figures 12 are related to the off-diagonal component $V_{zj}$ $(= V_{jz})$. Figures 12(a), (b) show the normalised voltage $V_w/V_S$ measured across the wire ends when the wire is excited by the external coil producing the longitudinal ac magnetic field. Without the dc current $I_b$ this characteristic is very small (it would be zero for an ideal

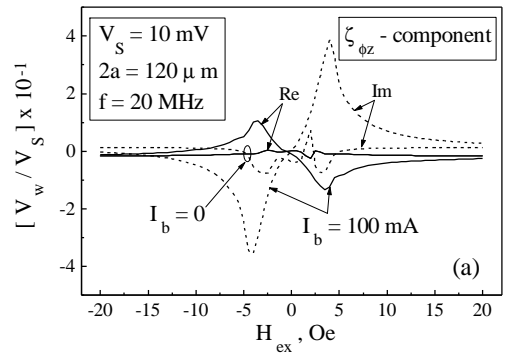

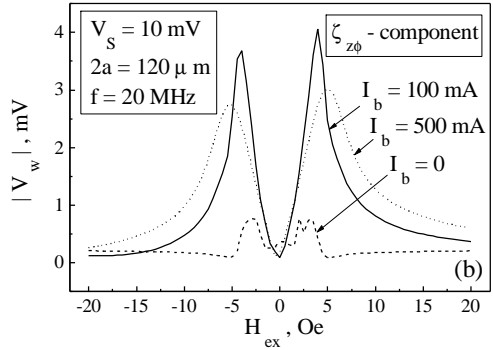

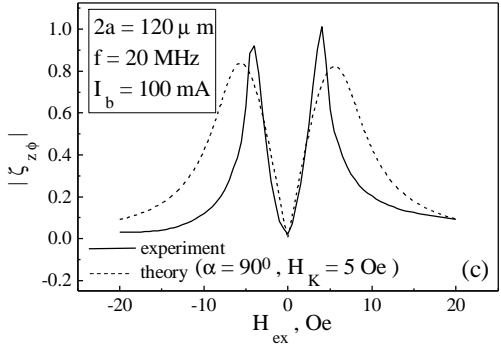

Fig. 12. Experimental plots of the off-diagonal impedance $V_{zj}(H_{ex})$ for different values of $I_b$. The result is presented in terms of the voltage ratio $V_w/V_S$ (which is proportional to $V_{zj}$): real and imaginary parts in (a), and the magnitude in (b). In (c), the impedance magnitude $|V_{zj}|$ vs. $H_{ex}$ (in values of $|V_w/V_s|$) is compared with the theoretical plots for a frequency of 20 MHz and $I_b = 100$ mA

circular domain structure since the averaged value $\sin\theta\cos\theta$ is zero) but it increases substantially when the current is enough to eliminate circular domains (compare the characteristics with $I_b = 0$ and $I_b = 100$ mA). Therefore, in the case of a circumferential anisotropy and a circular domain structure, the presence of $I_b$ is the necessary condition for the existence of the off-diagonal components of the impedance tensor. The off-diagonal component is anti-symmetrical with respect to the field $H_{ex}$, which is demonstrated in Fig. 12(a) by presenting both the real and imaginary parts. Such behaviour is an agreement with the theory (compare with Fig. 5(d)). A considerable increase in $I_b$ results in a decrease in sensitivity (see Fig. 12(b), $I_b = 500$ mA). In this case, the two opposite effects of $I_b$ are especially noticeable: (i) the transition to a single-domain structure (that increases $\varsigma_{z\varphi}$ and its field sensitivity), and (ii) the increase in the magnetic hardness in the circular direction (that decreases the sensitivity). Figure 12(c) shows the comparison of the experimental dependence with the calculated one at $I_b = 100$ mA. The experimental plot exhibits considerably faster decrease which may be related to some structural changes at the surface due to demagnetising effects since this component is very sensitive to the domain formation.

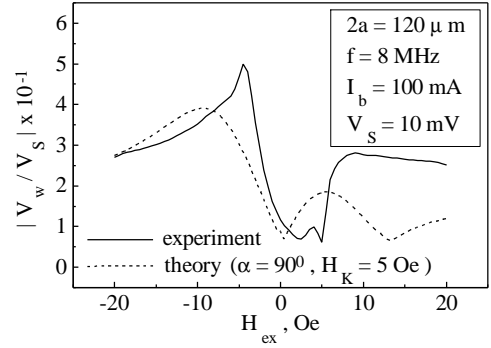

Fig. 13. Asymmetrical voltage response in the presence of the ac bias field. $I_b = 100$ mA. Theory and experiment.

Let us now suppose that a mixed excitation is used (Fig. 2(a)), when the wire is excited by both the ac current and the ac field $h_{ex}$ which is produced by the primary coil connected serially to the wire. The voltage measured across the wire is determined by equation (8) with $h_{ex} = 4\pi nLi/c$. In this case, the voltage $V_w$ involves both $V_{zz}$ and $V_{zj}$ components of the impedance tensor, combining symmetric and anti-symmetric terms with respect to $H_{ex}$. As a result, the voltage exhibits an asymmetric behaviour, even if the dc magnetic configuration does not have asymmetry, as shown in Fig. 13. In this case, the comparison with the theoretical result is more complicated. The coil gives an additional source of e.m.f which may cause the amplitude of the ac current to change during the experiment as well. [22,23]

**C. Helical anisotropy**

The case of helical anisotropy presents considerable interest since the effect of the dc current results in a completely different appearance of the field plots of the impedance. The experimental results reported here are obtained for the first time. A CoSiB amorphous wire has been studied, which has a relatively large anisotropy field of 8 Oe.

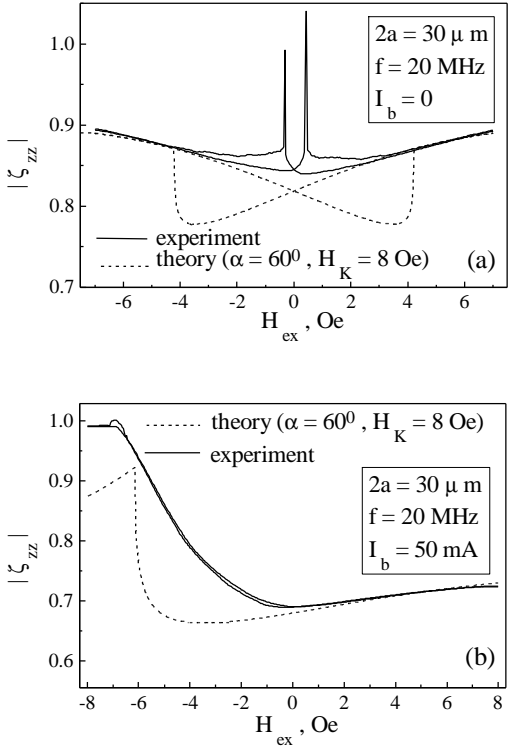

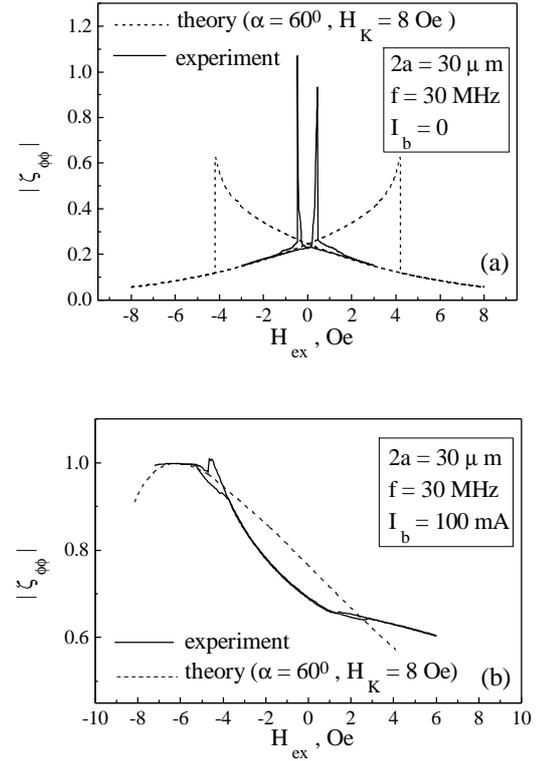

Fig. 14. Theoretical and experimental plots of $|V_{zz}|$ vs. $H_{ex}$ (in values of $|V_w/V_s|$) for a helical anisotropy ($\alpha = 60°$) for a frequency of 20 MHz. $I_b = 0$ in (a) and $I_b = 50$ mA in (b).

Fig. 15. Theoretical and experimental plots of $|V_{\varphi\varphi}|$ vs. $H_{ex}$ (in values of $|V_c/V_s|$) for a helical anisotropy ($\alpha = 60°$) for a frequency of 30 MHz. $I_b = 0$ in (a) and $I_b = 100$ mA in (b).

As it is known[40,41] it has a spontaneous helical anisotropy with the averaged angle of about 60°, which can be revealed by measuring the dc magnetisation loops in the presence of the dc current.[41] Figures 14 present the results for the longitudinal impedance $\varsigma_{zz}$. In this case the impedance exhibit a hysteresis. In Ref. 14, where the impedance of a wire with a twist induced helical anisotropy has been investigated, the hysteretic behaviour was not seen. In this case, the impedance field behaviour is related to domain wall permeability averaged over the ac magnetisation cycle due to irreversible helical-wall movement. The indication of irreversible non-linear processes involved is the considerable deviation from a sine-wave form of the measured voltage. The amplitude of the ac current exciting the wire used in Ref. 14 is 15 mA, which is sufficient to induce irreversible displacements of domain walls. In our experiment, such processes are not possible since $i_0 < 1$ mA. For $I_b = 0$, the experimental plot shows two sharp peaks at very small field corresponding to the coercivity field of the dc magnetisation process. The domain walls exist in this narrow field region and their linear dynamics gives a main contribution to the overall dynamic process. For fields larger than the coercivity, when the domain structure disappears, the impedance behaviour is determined by the ac magnetisation rotation. For these higher fields, there is a good agreement between the theory and the experiment. The theoretical jumps related to the irreversible rotational change in $M_0$ are not seen since in the experimental plot the dc magnetisation reversal is due to the domain processes. The effect of the dc current results in a gradual transition to non-hysteretic asymmetrical behaviour, shown in Fig. 14(b). The theoretical plot is in reasonable agreement with the experimental one. Certain discrepancy may be related to anisotropy dispersion, which is quite considerable in CoSiB amorphous wire.

Figures 15–16 present the field characteristics of $V_{\varphi j}$ and $V_{zj}$ components which change with the dc bias current in a characteristically similar manner. Note that $V_{zj}$ vs. $H_{ex}$ plot is very sensitive to the anisotropy angle. The theoretical curves describe two experimental maximuma very well for $\alpha = 60°$. This value of the anisotropy angle agrees with that found from the shift in the dc magnetisation loops.[41]

We can conclude that the theoretical model based on the single-domain magnetic structure agrees well with the numerous experimental data as far as the ac rotational magnetisation processes are responsible for the impedance change.

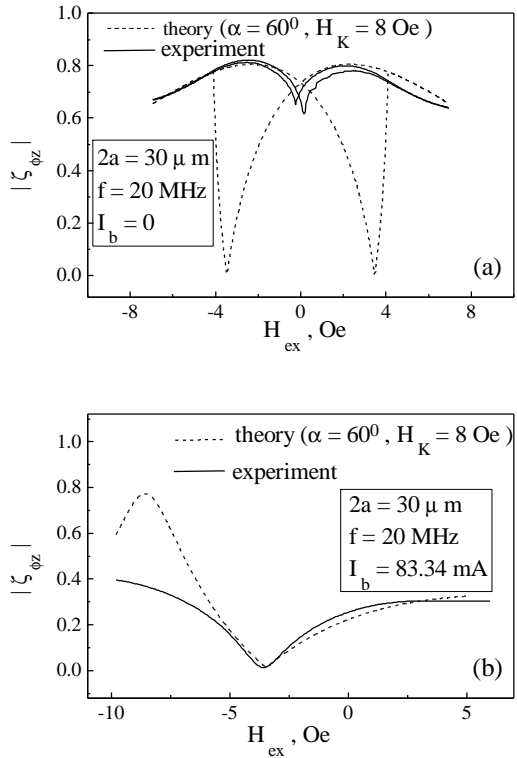

Fig. 16. Theoretical and experimental plots of $|V_{jz}|$ vs. $H_{ex}$ (in values of $|V_w/V_s|$) for a helical anisotropy ($\alpha = 60°$) for a frequency of 20 MHz. $I_b = 0$ in (a) and $I_b = 83.34$ mA in (b).

**Conclusion**

The surface impedance tensor approach has been used to study various types of MI characteristics in amorphous wires with a helical (circumferential) anisotropy. Regarding conceptual aspects of the MI effect, it has been demonstrated that a high sensitivity to the external field is caused by the dependence of the current density distribution on the static magnetic structure. Therefore, the characteristic field of the major impedance change is the anisotropy field, and the MI spectra are very broad (from a few MHz to hundreds of MHz for 30 μm diameter Co-based amorphous wire). Modifying the static magnetic structure, various types of the MI characteristics can be obtained: symmetrical or asymmetrical with respect to $H_{ex}$, without a hysteresis or exhibiting a hysteresis including of a bi-stable type. An interesting example is the change in MI characteristics in wires with a helical magnetic anisotropy under the effect of the dc current. Considering a tensor nature of the impedance, the use of the off-diagonal components results in asymmetrical MI in the presence of the ac bias, which is especially important for linear magnetic sensing.

The theory is based on the asymptotic-series expansion of the Maxwell equations. As far as the electrodynamic problems (such as the impedance analysis) are concerned, this method has been used here for the first time. It yields the analytical solution for the impedance tensor which is valid in entire frequency and magnetic field range (1 MHz – 1 GHz) of practical interest. The method has no restriction to a specific geometry. It can be expanded to consider practically important cases of 2-dimensional magnetic/metallic multilayers. The major limitation of the theory is considering a uniform magnetisation ignoring completely a radial distribution of permeability and the domain structure. Considering MI effects, the variation in permeability may not be important since the surface magnetisation gives predominant contribution. Regarding domain wall dynamics, it can be taken into account by modifying the permeability tensor on the basis of effective medium approximation for small field perturbations (Ref. 5,42). By this, the eddy currents due to the local wall displacements are averaged on the domain scale. Another restriction is the ignoring of exchange effects. This is accurate if the exchange length is smaller than the skin-depth, which is typically valid for frequencies under GHz-range.

The theoretical model has been tested comparing the results with the experimental data. In the case of helical anisotropy, the surface impedance tensor has been measured here for the first time. The theory agrees well with the numerical experimental data as far as the ac rotational magnetisation processes are responsible for the impedance change.